\documentclass[12pt]{article}

\usepackage{amsmath}
\usepackage{amssymb}
\usepackage{mathtext}
\usepackage{graphicx}

\begin{document}

\title{Generalized problem of two and four
Newtonian centers}
\author{Alexey V. Borisov, Ivan S. Mamaev\\
Institute of Computer Science, Udmurt State University\\
1, Universitetskaya str., 426034 Izhevsk, Russia\\
Phone/Fax: +7-3412-500295, E-mail: borisov@rcd.ru, mamaev@rcd.ru
}

\maketitle

\begin{abstract}
We consider integrable spherical analogue of the Darboux potential, which
appear in the problem (and its generalizations) of the planar motion of a
particle in the field of two and four fixed Newtonian centers. The obtained
results can be useful when constructing a theory of motion of satellites
in the field of an oblate spheroid in constant curvature spaces.
\end{abstract}


\newpage
\section{Introduction. }
The two-center problem is well known in classical celestial mechanics: two
fixed centers, with masses~$m_1^{}$ and~$m_2^{}$, attract some
``massless'' particle, moving in their field according to Newton's law.
The integrability of this problem was proved by Euler, by means of the separation
of variables~\cite{Ykobi}.

A qualitative analysis of the plane two-center problem was offered by
C.\,Char\-lier~\cite{Sharle} (see also~\cite{Deprit}); a qualitative
analysis of the spatial two-center problem can be found in the paper by
V.\,M.\,Alek\-se\-ev~\cite{Alekseev}. Note also that it was Lagrange who
observed that the two-center problem remains integrable upon addition of
the potential of an elastic spring attached to the midpoint of the
rectilinear segment connecting both centers. Lagrange also studied the
limiting case of the problem, where one of the two-centers and its mass
tend to infinity. In the limit we have the problem of a point subject to
the superposition of the field of a Newtonian center (the Kepler problem)
and a homogeneous field. The corresponding separation of variables and a
qualitative study of this problem was done by M.\,Born in his book on
atomic mechanics~\cite{111}, in connection with the splitting of a hydrogen
atom's spectral lines observed when it is put into an
electric field (the Stark effect).

A more general two-dimensional integrable system, which incorporates the
two-center problem, was found by G.\,Darboux~(1901)~\cite{Darboux}, by
means of the method of the separation of variables. In this paper Darboux also
obtained the existence conditions for an additional quadratic integral in
case of a natural system on a plane. Later, these conditions were
described by Whittaker~\cite{Uiteker}.

Consider a particle of unit mass, moving on~$\mathbb{R}^2 = \{x,y\}$ in
the potential field
\begin{equation}
\label{e1}
V = \frac{A}{x^2} + \frac{A'}{y^2} +
\frac{B}{r} + \frac{B'}{r'} + \frac{B_1^{}}{r_1^{}} + \frac{B_1'}{r_1'} +
C \rho^2,
\end{equation}
where $A,\,A',\,B,\,B',\,B_1^{},\,B_1',\,C = \mathrm{const}$.
Here,~$r$ and~$r'$ are the real distances between~$m$ and the two
identical real centers positioned at the points~$(-c,\,0)$, $(c,\,0)$ on
the abscissa axis, $r=\sqrt{(x-c)^2+y^2}$, $r'=\sqrt{(x+c)^2+y^2}$; $\rho$
is the distance between~$m$ and the origin~$O$; $r_1^{}$ and~$r_1'$ are
the ``complex distances'' between~$m$ and the two imaginary
centers~$(0,\,di)$ and~$(0,\,-di)$, $r_1^{}=\sqrt{x^2+(y-id)^2}$,
$r_1'=\sqrt{x^2+(y+id)^2}$ (see Fig.~1).


For the potential~\eqref{e1} to be real, it is necessary that $B_1'$
and~$B_1$ are complex conjugate: $\overline{B_1}'=B_1$. As is shown
in~\cite{Darboux}, if~$d = c$, the system~\eqref{e1}
allows the separation of variables in elliptic coordinates
$$
x = c\,\mathrm{ch} v \cos u, \qquad y = c\,\mathrm{sh} v \sin u
$$
and has an additional first
integral, which is quadratic in the momenta.

Let us examine some special cases of the potential~\eqref{e1}. The case
of~$B_1^{} = B_1'= 0$ was studied by G.\,Liouville (a more
special case of $A = A' = B_1^{} = B_1' = 0$ was, as we already mentioned,
described by Lagrange).

It was shown in the paper~\cite{Aksenov} that the problem of a particle
moving in the field of two complex conjugate centers, i.\,e., when~${A = A'
= B = B' = C = 0}$ in~\eqref{e1}, is integrable in the three-dimensional space
and makes a good approximation to the problem of a satellite's motion in
the field of an oblate spheroid (eg., the motion of an artificial satellite of the
Earth).

In the paper by I.\,S.\,Kozlov~\cite{Kozlov}, the problem of the plane
motion of a particle in the field of four fixed centers (two real and two
complex) was integrated in terms of quadratures and further studied.
Besides, in~\cite{Kozlov}, several interpretations of this problem were
offered, with reference to actual problems of applied celestial mechanics.

\section{The Kepler problem. The two-center problem on a sphere and a pseudosphere.
Historical notes}

Systematic generalization of various problems from classical and celestial
mechanics to constant curvature spaces (a three-dimensional sphere~$S^3$, as
well as a pseudo-sphere~$L^3$, or Lobachevsky space) was done by W.\,Killing
in his extensive, but, unfortunately, almost forgotten paper~\cite{Kill2}.

Note also that beside Killing, in the~19th century, non-Euclidean
mechanics in constant curvature spaces was studied by R.\,Lipschitz,
F.\,Schering and H.\,Liebmann. It is interesting that though a whole chapter
from Liebmann's textbook on non-Euclidean geometry~\cite{Liebmann}
concerned the generalization of Newton's law of attraction, study of the
Kepler problem, and reformulation of Kepler's laws for the cases of a sphere
and a pseudosphere, similar results were independently and almost
simultaneously rediscovered in the 20th century by several
authors~\cite{KHarin2,Chernikov2,Higgs2,Kozlov24_2,Verpry,Ibeda,Slaw,Chernikov}.
The classical paper by E.\,Schr\"odinger~\cite{Sch_2} should also be
mentioned, where he studied a quantum analog of the Kepler problem in a
curved space, implicitly assuming that the corresponding classical problem
was integrable. An analogue of Newton's law of attraction for~$L^3$ was
known to J.\,Bolyai, and~N.\,I.\,Lobachevsky, and  for~$S^3$~--- to P.\,Serret.

W.\,Killing in~\cite{Kill2} studied, among other things, the problems of
$n$-dimensional dynamics in constant curvature spaces, including the
dynamics of an~$n$-dimensional rigid body. An up-to-date analysis can be
found in~\cite{Dombrow} (see also~\cite{Dubna1}).

The generalization of the two-center problem to constant curvature spaces
is also due to W.\,Killing, who integrated this problem, using the method
of the separation of variables. It was independently solved in~\cite{KHarin2},
where a more general problem was studied, similarly to what Lagrange did
by introducing an elastic interaction potential in the plane two-center
problem. In the papers~\cite{Voz,Vozmi}, a bifurcational analysis of the
two-center problem on a sphere and on the Lobachevsky plane was offered.
In~\cite{Dubna1} we examined the spatial two-center problem from the
standpoint of its reduction and integrability; we also studied other
integrable and non-integrable problems of celestial mechanics in curved
spaces (including the restricted two- and three-body problems, behaviour of
libration points, dynamics of rigid bodies).

In this paper, we will offer an explicit algebraic expression for the
first integral of the generalized two-center problem
from~\cite{Kill2,KHarin2}, and show a new analogue of the problem of four
Newtonian centers and~$n$ Hookian centers. In this paper, we study only
the case of a two-dimensional sphere~$S^2$, though all the reasoning can
easily be extended to a pseudosphere~$L^2$. Certain (not all) results are
generalized to a three-dimensional sphere~$S^3$ (or a pseudosphere~$L^3$).

\section{Generalization of the two-center problem to~$S^2$. Additional
quadratic integral}

Assume that a unit sphere~$S^2$ is given in the three-dimensional
space~$\mathbb{R}^3 = \{q_1^{},\,q_2^{},\,q_3^{}\}$ by~$|\mathbf{q}|^2 = q_1^2 +
q_2^2 + q_3^2 = 1$ and denote by~$\mathbf{q}=(q_1^{},\,q_2^{},\,q_3^{})$, $\mathbf{p} =
(p_1^{},\,p_2^{},\,p_3^{})$ the redundant coordinates and momenta,
respectively. Now if we introduce the angular momentum vector~$\mathbf{M} = \mathbf{p}\times
\mathbf{q}$ and put~$\boldsymbol\gamma = \mathbf{q}$, it is easy to show~\cite{Dubna1,Bo,118}
that the equations of motion in an arbitrary potential~$V = V(\mathbf{q}) =
V(\boldsymbol\gamma)$ can be presented as a Hamiltonian system with the Poisson
bracket defined by the algebra~$e(3) = so(3) \oplus_s^{} \mathbb{R}^3$:
\begin{equation}
\label{e2}
 \{M_i^{},\,M_j^{}\} = \varepsilon_{ijk}^{}
M_k^{}, \qquad \{M_i^{},\,M_j^{}\} = \varepsilon_{ijk}^{}M_k^{}, \qquad
\{\gamma_i^{},\,\gamma_j^{}\} = 0
\end{equation}
and the
Hamiltonian
\begin{equation}
\label{e3}
H = \frac12 (\mathbf{M},\,\mathbf{M}) + V (\boldsymbol\gamma).
\end{equation}
From~\eqref{e2}, \eqref{e3} we obtain
the equations
$$
\dot{\mathbf{M}} = \boldsymbol\gamma \times
\frac{\partial{V}}{\partial{\boldsymbol\gamma}}, \dot{\boldsymbol\gamma} = \boldsymbol\gamma \times \bf{M},
$$
which coincide with the equations of motion of a
spherical top in the potential~$V(\boldsymbol\gamma)$~\cite{Dubna1,Bo}.

The bracket \eqref{e2} is degenerate and has two Casimir
functions:~$F_1^{} = (\mathbf{M},\,\boldsymbol\gamma)$, $F_2^{}= (\boldsymbol\gamma,\,\boldsymbol\gamma)= 1$.
For the problem of a point moving on a sphere, it is
necessary that~$F_1^{} = (\mathbf{M},\,\boldsymbol\gamma) = (\mathbf{p}\times \boldsymbol\gamma,\,\boldsymbol\gamma) = 0$.

It is well known that the analogues of the Newtonian and Hookian potentials
on~$S^2$ are, respectively,~$U_1^{} = \mu \cot \theta$ and~$U_2^{} = c\tan^2
\theta$, $\mu,\,c = \mathrm{const}$, where~$\theta$ is measured from a certain fixed
pole on the sphere~\cite{Kill2,KHarin2}.

Consider the potential
\begin{equation}
\label{e4}
 V = - \mu_1^{} \cot\theta_1^{} - \mu_2^{}
\cot\theta_2^{},
\end{equation}
 where $\mu_1^{}$, $\mu_2^{}$ are the intensities of the
Newtonian centers, while~$\theta_i^{}$ is the angle between the radius-vector
of a particle and the radius-vector of the~$i$-th center.
Place the
Newtonian centers at the points~$\mathbf{r}_1^{} = (0,\,\alpha,\,\beta)$, $\mathbf{r}_2^{} =
(0,\,-\alpha,\,\beta)$, $\alpha^2 + \beta^2 = 1$, and add to~\eqref{e4}, for the
sake of generality, the potentials of three Hookian centers located at the
mutually perpendicular axes~$\frac{1}{2} \sum c_i^{}/ \gamma_i^2$
($c_i^{} = \mathrm{const}$).
We should also introduce the quadratic
potential~$C(\alpha^2 \gamma_2^2 - \beta^2 \gamma_3^2)$ $C \neq 0$, which is a
special case of Neumann's potential. On the level~$(\mathbf{M},\,\boldsymbol\gamma) =
0$ we find two commuting functions~$\{H,\,F\} = 0$, quadratic
in~$\mathbf{M}$~\cite{118,Mamaev}:
\begin{equation}
\label{e5}
\begin{split}
H & = \frac12 \mathbf{M}^2 - \mu_1^{}
\frac{\beta \gamma_3^{} + \alpha \gamma_2^{}} {\sqrt{\gamma_1^2 + \beta_1^{}
\gamma_2^2 + \alpha^2 \gamma_3^2 - 2 \alpha \beta \gamma_2^{}
\gamma_3^{}}}-\\
\,& - \mu_2^{}\frac{\beta \gamma_3^{} - \alpha \gamma_2^{}}
{\sqrt{\gamma_1^2 + \beta_1^{} \gamma_2^2 + \alpha^2 \gamma_3^2
+ 2 \alpha \beta \gamma_2^{} \gamma_3^{}}} + \frac12 c_1^{} \frac{\gamma_2^2 +
\gamma_3^2}{\gamma_1^2}+\\
\,& + \frac12 c_2^{}\frac{\gamma_1^2+ \gamma_3^2}{\gamma_2^2}+ \frac12
c_3^{} \frac{\gamma_1^2 + \gamma_2^2}{\gamma_3^2} + C(\alpha^2 \gamma_2^2 -
\beta^2\gamma_3^2),\\
F & = \alpha^2 M_2^2 - \beta^2 M_3^2 + 2 \alpha \beta(V_1^{} - V_2^{}) - \\
\,& - \frac{c_1^{}}{\gamma_1^2}(\beta^2 \gamma_2^2 - \alpha^2 \gamma_3^2) -
\frac{c_2^{}}{\gamma_2^2} \beta^2 \gamma_1^2 + \frac{c_3^{}}{\gamma_3^2}
\alpha^2 \gamma_1^2 + 2 C \alpha^2 \beta^2 \gamma_1^2,
\end{split}
\end{equation}
where $\mu_1^{}$,
$\mu_2^{}$, $\alpha$, $\beta$, $c_1^{}$, $c_2^{}$, $c_3^{}$, $C = \mathrm{const}$,
and the functions~$V_1^{}$, $V_2^{}$ are:
\begin{equation}
\begin{split}
\label{e6}
 V_1^{} & =
\frac{\mu_1^{}(\beta \gamma_2^{} + \alpha \gamma_3^{})} {\sqrt{\gamma_1^2 +
\beta^2 \gamma_2^2 + \alpha^2 \gamma_3^2 - 2 \alpha \beta
\gamma_2^{} \gamma_3^{}}},\\
V_2^{} & = \frac{\mu_2^{}(\beta \gamma_2^{} - \alpha \gamma_3^{})}
{\sqrt{\gamma_1^2 + \beta^2 \gamma_2^2 + \alpha^2 \gamma_3^2 + 2 \alpha \beta
\gamma_2^{} \gamma_3^{}}}.
\end{split}
\end{equation}
 The function~$H$ is the Hamiltonian, and~$F$
is an additional quadratic integral. As it is noted in~\cite{Dubna1}, the
integrability of the corresponding three-dimensional ($S^3$) problem
closely depends on whether a Hookian center (with
potential~$c/\gamma_3^2$) can be added (at some point of the arc joining
the Newtonian centers) to the two-center problem~\eqref{e4} without
violating the problem's integrability. Indeed, the
term~$c/\gamma_3^2$, $c = \mathrm{const}$ appears in the three-dimensional
case as a result of the Routh reduction procedure, which uses the cyclic
integral. This integral is due to the equations' invariance under
rotations (group~$SO(2)$), in the plane perpendicular to the plane of the
two centers.

The system \eqref{e5} is of the Liouville type and can be integrated in
sphero-conical coordinates~$u_1^{}$, $u_2^{}$, ($0 < u_1^{} < \alpha$, $0 <
u_2^{} < \beta$) given by
\begin{equation}
\begin{split}
\label{e7}
\gamma_1^{} & = \sqrt{u_1^{}u_2^{}}/(\alpha \beta),\\
\gamma_2^{} & = \sqrt{(\alpha^2 - u_1^{})(\alpha^2 + u_2^{})}/\alpha,\\
\gamma_3^{} & = \sqrt{(\beta^2 + u_1^{})(\beta^2 - u_2^{})}/\beta.
\end{split}
\end{equation}
Note, however, that obtaining the integrals \eqref{e5} in the algebraic
form is quite a non-trivial problem, as its solution implies dealing with
an inverse sphero-conical transformation.

As A.\,Albouy informed us, the two-center problem on~$S^2$ (or~$L^2$) can
be transformed to the traditional Euler problem of two centers by means of
the central (gnomonic) projection and a suitable transformation of time.
However, we cannot yet prove this statement.

\section{The problem of four Newtonian centers on~$S^2$}

Consider the potential on a sphere:
\begin{multline}
\label{e8}
V_{\text{Im}}^{} = \xi_1^{} \cot \theta_1^{} + \xi_2^{} \cot \theta_2^{} = \\
=\xi_1^{} \frac{\mu \gamma_1^{} {+} i \nu \gamma_3^{}}{\sqrt{(\mu^2 -
\nu^2)^2 - (\mu \gamma_1^{} + i\nu \gamma_3^{})^2}} {+} \xi_2^{} \frac{\mu
\gamma_1^{} - i\nu \gamma_3^{}} {\sqrt{(\mu^2-\nu^2)^2 - (\mu \gamma_1^{}
- i\nu \gamma_3^{})^2}},
\end{multline}
where $\mu^2 - \nu^2 = 1$, $\xi_1^{},\,\xi_2^{} = \mathrm{const}$.

This potential corresponds to the two-center problem on sphere. The
intensities of the centers are ``complex'' and the centers themselves are
equidistant (complex conjugate) from the pole (Fig.~2). For the potential
to be real, it is necessary that~$\overline{\xi}_1^{} = \xi_2^{}$. As in the
Euclidean case, the potential~\eqref{e8} can be regarded as a certain
approximation to the problem of a particle moving in the field of an
oblate spheroid in a curved space.


The system with the potential \eqref{e8} is also separable in the
sphero-conical coordinates~\eqref{e7}, provided that
\begin{equation}
\label{e9}
\mu =
\frac{\beta}{1-\alpha^2}, \qquad \nu = \frac{\alpha \beta}{1-\alpha^2}.
\end{equation}

In the coordinates \eqref{e7}, the potential~$V + V_{\text{Im}}^{}$ is also
separable. This potential (for~$c_i^{}=0$) corresponds to the problem of
four fixed centers, two imaginary and two real, which belong to two
mutually perpendicular planes through the pole (see Fig.~2). Here, as in
the planar case, when the distance between the real centers is fixed, the
distance between the complex centers is also not arbitrary: it is uniquely
defined by~\eqref{e9}.

It is easy to show that the potentials
\begin{equation}
\label{e10}
V_G^{} =
\frac12\left(\sum{c_i^{}}/{\gamma_i^2}\right)\!, \qquad  V_N^{} = C(\alpha^2
\gamma_2^2 - \beta^2 \gamma_3^2) \qquad c_i^{},\,C = \mathrm{const}.
\end{equation}
can be added
(without violating integrability) to the four centers problem and it
results in a more general system, separable in the coordinates~\eqref{e7}.
The potentials~$V,\,V_{\text{Im}}^{},\,V_G^{},\,V_N^{}$ written in terms of the
variables~\eqref{e7} look like:
$$
\begin{gathered}
V = \frac{(\mu_1^{} + \mu_2^{})\sqrt{(\alpha^2 - u_1^{})
(\beta^2 + u_1^{})} + (\mu_1^{} - \mu_2^{})\sqrt{(\alpha^2 + u_2^{})
(\beta^2 - u_2^{})}}{u_1^{} + u_2^{}},\\
V_{\text{Im}}^{} = \frac{(\xi_1^{} + \xi_2^{})
\sqrt{u_2^{}(\beta^2 - u_2^{})} + i(\xi_1^{} -
\xi_2^{})\sqrt{u_1^{}(\beta^2 + u_1^{})}}{u_1^{} + u_2^{}},\\
\frac{1}{\gamma_1^2} = \beta^2 \frac{(\beta^2 - u_2^{})^{-1} - (\beta^2 + u_1^{})^{-1}}
{u_1^{} + u_2^{}},\\
\frac{1}{\gamma_2^2} = \alpha^2 \frac{(\alpha^2 - u_1^{})^{-1} - (\alpha^2 + u_2^{})^{-1}}
{u_1^{} + u_2^{}},\\
\frac{1}{\gamma_3^2} = \alpha \beta \frac{u_1^{-1} + u_2^{-1}}{u_1^{} +
u_2^{}},
V_N^{} =  C\frac{u_1^2 - u_2^2}{u_1^{} + u_2^{}}.
\end{gathered}
$$
One can easily show that in the limit~${R \to\infty}$ (the case of
Euclidean plane) the total potential~$V + V_{\text{Im}}^{} + V_G^{} + V_N^{}$
becomes the Darboux potential~\eqref{e1}. Note that this potential, or
even~$V + V_{\text{Im}}^{}$, can no longer be generalized to the corresponding
integrable potential of the three-dimensional problem ($S^3$), because
there is no cyclic integral, though, taken individually, the
potentials~$V$ and~$V_{\text{Im}}$ allow such a generalization.


\section{The problem of $n$ Hookian centers on a sphere}

Let us present one more integrable modification of the problem of a mass
point moving in the field of the Hookian potentials~${c_i^{}}/{({\boldsymbol\gamma}
,\,\mathbf r_i^{})^2}$, $c_i^{} = \mathrm{const}$, where the Hookian centers of
attraction~$\mathbf r_i^{}$, $i = 1,\,2,\ldots, n$ do not belong to mutually
orthogonal axes, but are placed arbitrarily on an equator~\cite{Mamaev}
(Fig. 3).

When $(\mathbf{M},\,{\boldsymbol\gamma})=0$, the Hamiltonian and the additional integral
are
\begin{equation}
\begin{split}
\label{e11}
H = & \frac12 \mathbf{M}^2 + \frac12 \sum_{i=1}^n
 \frac{c_i^{}}{(\mathbf{r}_i^{},\,\boldsymbol\gamma)^2} + U(\gamma_3^{})\\
F = & M_3^2 + (1 - \gamma_3^2)\sum_{i=1}^n \frac{c_i^{}}{(\mathbf{r}_i^{},\,\boldsymbol\gamma)^2}.
\end{split}
\end{equation}
 There is an arbitrary function~$U(\gamma_3^{})$
in~\eqref{e11}. This function means addition of an arbitrary ``central''
field with the center on a perpendicular to the plane of the Hookian
potentials (Fig.~3). For example, one more Hookian center can be placed at
the pole. This implies (see~\cite{118}) that the spatial problem of a
point moving on a three dimensional sphere~$S^3$ under the action of~$n$
Hookian centers on its equator is also integrable.

Note that a Euclidean analogue of the problem in question is trivial, as
even in Cartesian coordinates it yields~$n$ linear oscillators. In this
case, the Hookian centers can be arbitrarily scattered in~$\mathbb{R}^2$.
In the case of curved space, even on a two-dimensional sphere, the problem
of motion in the field of three arbitrarily placed Hookian centers is not
integrable, as simulations reveal chaos in this system. The quadratic
integral~$F$ in~\eqref{e11} is due to the fact that the problem is
separable in the spherical coordinates~$(\theta,\,\varphi)$. Indeed, the
Hamiltonian~$H$ can be written as
\begin{multline}
\label{e12}
H = \frac12\left(p_\theta^2 + \frac{p_\varphi^2}{\sin^2 \theta}\right) +
\frac12\sum_{i=1}^n \frac{c_i^{}}{\sin^2 \theta \cos^2 (\varphi - \varphi_i^{})} +
U(\theta) = \\
 = \frac12 p_\theta^2 + \frac{1}{\sin^2 \theta}\left[p_\varphi^2 +
\sum_{i=1}^n \frac{c_i^{}}{\cos^2 (\varphi - \varphi_i^{})}\right] +
U(\theta),
\end{multline}
where $\theta$, $\varphi$ are the coordinates of the moving mass point,
while~$\varphi_i^{}$ defines the position of the~$i$-th Hookian center on
the equator (Fig.~3). The expression in square brackets is an additional
integral of motion~\eqref{e11}.

\section{Acknowledgements}

The authors thank A.\,Albouy and V.\,V.\,Kozlov for useful discussions.
This work was supported from the program ``State Support for Leading
Scientific Schools'' (136.2003.1); additional support was given
by the Russian Foundation for Basic Research (04-05-64367), the
U.S. Civilian Research and Development Foundation (RU-M1-2583-MO-04)
and the INTAS (04-80-7297).

\begin{figure}[!ht]
\begin{center}
\includegraphics{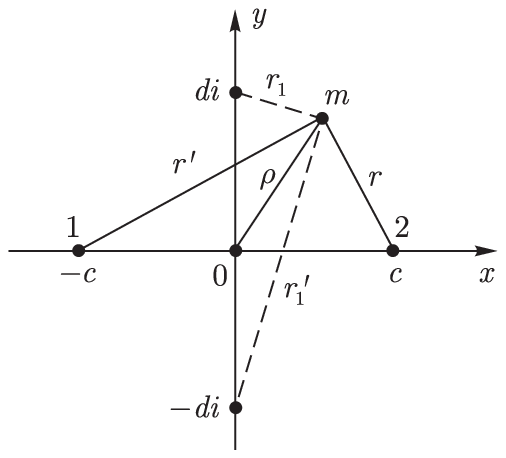}
\end{center}
\caption{Location of real and ``imaginary'' centers on a plane}
\end{figure}

\begin{figure}[!ht]
\begin{center}
\includegraphics{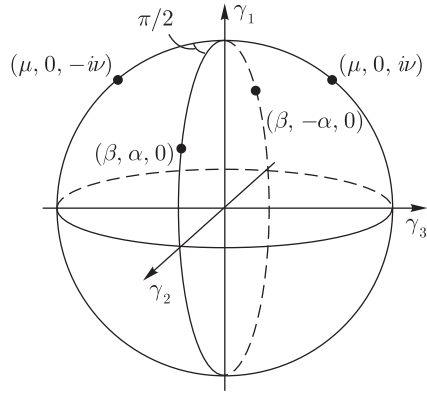}
\end{center}
\caption{Location of real and ``imaginary'' centers on a sphere}
\end{figure}

\begin{figure}[!ht]
\begin{center}
\includegraphics{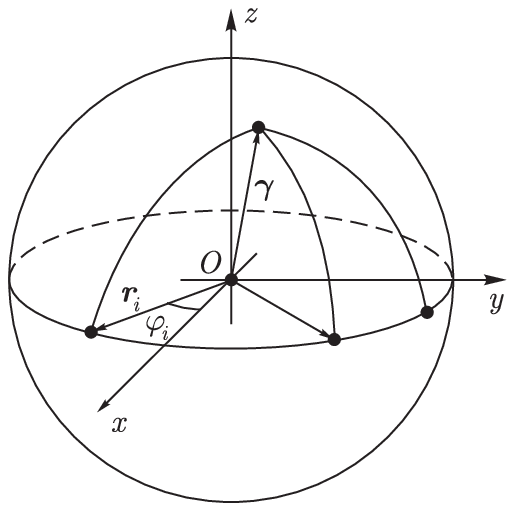}
\end{center}
\caption{Mutual location of a particle and Hookian centers on a sphere}
\end{figure}


\begin{thebibliography}{99}

\bibitem{Aksenov}
Aksenov, Ye.\,P., Grebenikov, Ye.\,A. and Demin, V.\,T.: 1963, `Generalized problem
of two fixed centers and its application in the theory of motion of the
Earth's artificial satellites', \emph{Astron. J.} \textbf{40}, 363--372.

\bibitem{Alekseev}
Alekseev, V.\,M.: 1665, `Generalized spatial problem of two fixed centers.
Classification of motions', \emph{ITA Bulletin} \textbf{10}, 241--271.

\bibitem{Bo}
Bogoyavlensky, O.\,I.: 1985, `Integrable cases of rigid body dynamics and
integrable systems on spheres~${S}^n$', \emph{Izv. AN SSSR} \textbf{49}, 899--915.

\bibitem{118}
Borisov, A.\,V. and Mamaev, I.\,S.: 2001, \emph{Rigid Body Dynamics,} Izhevsk, SPC
``RCD''. (In Russian)

\bibitem{Dubna1}
Borisov, A.\,V. and Mamaev, I.\,S.: 1999,  \emph{Poisson structures and Lie algebras
in Hamiltonian mechanics,}  Izhevsk, SPC ``RCD''.
(In Russian)

\bibitem{111}
Born, M.: 1925,  \emph{Vorlesungen \"{u}ber Atommechanik,} Berlin, Springer.

\bibitem{Sharle}
Charlier, C.\,L.: 1927,  \emph{Die Mechanik des Himmels,} Berlin, Walter de Gruyter \& Co.

\bibitem{Chernikov2}
Chernikov, N.\,A.: 1992, `The Kepler problem in the Lobachevsky space and its
solution', \emph{Acta Phys. Polonica} \textbf{23}, 115--119.

\bibitem{Chernikov}
Chernikov, N.\,A.: 1992, `The relativistic Kepler problem in the Lobachevsky
space',  \emph{Acta Phys. Polonica} \textbf{24}, 927--950.

\bibitem{Darboux}
Darboux, G.: 1901, `Sur un probl\'{e}me de m\'{e}canique', \emph{Archives
N\'{e}erlandaises de Sciences} \textbf{6}, 371--376.

\bibitem{Deprit}
Deprit, A.: 1962, `Le probl\'{e}me des deux centers fixes', \emph{Bull. Nath. Belg.}
\textbf{142}, 12--45.

\bibitem{Dombrow}
Dombrowski, P., Zitterbarth, J.: 1991, `On the planetary motion in the $3$
standard spaces $M_k^3$ of constant curvature $k \in \mathbb{R}$',
\emph{Demonstratio Math.} \textbf{24}, 375--458.

\bibitem{Higgs2}
Higgs, P.\,W.: 1979, `Dynamical symmetries in a spherical geometry. I.',
\emph{J.~Phys.~A.} \textbf{12}, 309--323.

\bibitem{Ibeda}
Ikeda, M., Katayama, N.: 1982, `On generalization of Bertrand's theorem to
spaces of constant curvature', Tensor N.\,S. \textbf{38}, 37--40.

\bibitem{Ykobi}
Jacobi, C.\,J.: 1884, \emph{Vorlesungen \"{u}ber Dynamik}, Aufl. 2, Berlin, G.
Reimer.

\bibitem{Kill2}
Killing, W.: 1885, `Die Mechanik in den Nicht-Euklidischen Raumformen',
\emph{J.~Reine Angew. Math.} \textbf{98}, 1--48.

\bibitem{Kozlov}
Kozlov, I.\,S.: 1974, `The problem of four fixed centers and its applications
to the theory of motion of celestial bodies', \emph{Astron. J.} \textbf{51},
191--198.

\bibitem{Kozlov24_2}
Kozlov, V.\,V.: 1994, `On dynamics in constant curvature spaces', \emph{Vestnik MGU, Ser. Math. Mech.}
28--35.

\bibitem{KHarin2}
Kozlov, V.\,V., Harin, A.\,O.: 1992, `Kepler's problem in constant curvature
spaces', \emph{Cel. Mech. and Dyn. Astr.} \textbf{54}, 393--399.

\bibitem{Liebmann}
Liebmann, H.: 1905, \emph{Nichteuklidische Geometrie}, Leipzig,
G.\,J.\,G\"{o}schen'sche Verlagshandlung.

\bibitem{Mamaev}
Mamaev I.\,S.: 2003, `Two integrable systems on a two-dimensional sphere',
Doklady RAN \textbf{389}, 338--340.

\bibitem{Sch_2}
Schr\"odinger, E.: 1940, `A method of determining of quantum-mechanical
eigenvalues and eigenfunctions', \emph{Proceedings of the Royal Irish Academy}
\textbf{46A}, 9--16.

\bibitem{Slaw}
Slawianowski, J.: 1980, `Bertrand systems on $so(3,{R})$, $su(2)$', \emph{Bull. de
l'Academie Polonica des Sciences} \textbf{28}, 83--94.

\bibitem{Verpry}
Velpry, C.: 2000, `Kepler laws and gravitation in non-Euclidean
{\rm(}classical\/{\rm)} mechanics', \emph{Heavy Ion Physics} \textbf{11}, 131--146.

\bibitem{Vozmi}
Vozmischeva, T.\,G.: 2000, `Classification of motions for generalization of
the two center problem on a sphere', \emph{Cel. Mech. and  Dyn. Astr.} \textbf{77}, 37--48.

\bibitem{Voz}
Vozmischeva, T.\,G., Oshemkov, A.\,A.: 2002, `Topological analysis of the
two-center problem on two-dimensional sphere', \emph{Math. Sbornik} \textbf{193}, 3--38.

\bibitem{Uiteker}
Whittaker, E.\,T.: 1959, \emph{Analytical Dynamics,} Cambridge
University Press.

\end{thebibliography}
\end{document}